\documentclass[prd,tightenlines,nofootinbib,amsfonts,amssymb,amsmath,color,english,12pt]{article}
\usepackage{atbegshi}
\AtBeginDocument{\AtBeginShipoutNext{\AtBeginShipoutDiscard}}
\usepackage[affil-it]{authblk}
\usepackage[T1]{fontenc}
\usepackage[latin9]{inputenc}
\usepackage{amsmath}
\usepackage{babel}
\usepackage{amsmath}
\usepackage{amssymb}
\usepackage{hyperref}
\usepackage{geometry}
\usepackage{color}
\usepackage{graphicx}
\usepackage{subfig}
\usepackage[font=small,labelfont=bf,margin=1.2cm]{caption}
\usepackage{verbatim}

 \global\long\def\halfsize{0.45\columnwidth}

\begin{document}
\title{\bf Static cosmological solutions in quadratic gravity}
\author[1]{Daniel M\"uller} 
\thanks{muller@fis.unb.br}
\author[2,3]{Alexey Toporensky \thanks{atopor@rambler.ru}}
\affil[1]{Instituto de F\'{i}sica, Universidade de Bras\'{i}lia, Caixa Postal
04455, 70919-970 Bras\'{i}lia, Brazil}
\affil[2]{Kazan Federal University, Kazan 420008, Republic of Tatarstan, Russian Federation}
\affil[3]{Sternberg Astronomical Institute, Moscow University, Moscow 119991, Russian Federation}
\date{\today}
\maketitle
\abstract{{\small
We consider conditions for existence and stability of a static cosmological solution in quadratic gravity. It appears 
 that such a solution for a Universe filled by only one type of perfect fluid is possible in a wide range
of the equation of state parameter $w$ and for both positively and negatively spatially curved Universe.
We show that the static solution for the negative curvature is always unstable if we require positive
energy density of the matter content. On the other hand, a static solution with positive spatial curvature
can be stable under certain restrictions. Stability of this solution with respect to isotropic perturbation
requires that the coupling constant with the $R^2$ therm in the Lagrangian of the theory  is positive, 
and the equations of state parameter $w$ is located in a rather narrow interval. Nevertheless, the stability 
condition does not require violation of the Strong Energy Condition. Taking into account anisotropic perturbations
leads  to further restrictions on the values of coupling constants and the parameter $w$. 

}}
\section{Introduction}
Cosmological dynamics in modified gravity is much richer than in GR. Already in $R+R^2$ theory a possibility to
get acceleration expansion of the Universe without any special matter violating the Strong Energy Conditions appears.
This leads to Starobinsky inflation \cite{Starobinsky:1980te}, being the first and still one of the most popular and viable inflation scenarios \cite{Gorbunov:2014ewa}, \cite{Mishra:2018dtg}  \cite{Mishra:2019ymr}.
There are other modifications of cosmological dynamics, such as existing of a stable  isotropic past attractor for Bianchi I
Universe  \cite{Barrow:2006xb}, \cite{Netto_2016},  \cite{Castardelli_dos_Reis_2019}, \cite{Muller:2017nxg}.

Quadratic gravity is not something so new, as Weyl and others investigated even in 1918 \cite{weyl1918gravitation}. During the 60's there's the pioneering work of Buchdahl \cite{buchdahl1962gravitational}. In the context of Schwinger approach it seems that Utiyama together with de Witt were the first to obtain the first loop corrections to the classical Einstein action \cite{Utiyama:1962sn}. For a historical review see, for instance \cite{schmidt2007fourth}.

In the present paper we consider another interesting type of solutions, which are very problematic in GR -- stable 
static solutions. The story of static solutions goes back to Einstein in 1917 and his famous static positively curved Universe
filled by dust with the energy density $\rho$ and cosmological constant $\Lambda$ so that $\Lambda=\rho/2$ \cite{Einstein:1917ce}. It can be easily generalized to the case of a arbitrary matter equation
of state $w_1$ unless $w_1<-1/3$ and has the form $\Lambda=\rho(1+3w_1)/2$ \cite{PhysRevD.76.084005}. Moreover, the cosmological constant can be replaced by any matter with $w_2<-1/3$ with the following simple condition $\rho_1(1+3w_1)+\rho_2(1+3w_2)=0$.
However, it is known that if we require positiveness of the energy density of these types of matter, the static
solution is unstable as already remarked by Eddington in 1930 in the case of one fluid only \cite{10.1093/mnras/90.7.668}. Stability was further investigated for strict GR for example by \cite{Harrison:1967zza}, \cite{Gibbons:1987jt}, \cite{Gibbons:1988bm} and \cite{Barrow:2003ni}. It turns out that although Einstein static Universe is unstable for homogeneous oscillations it is stable for general inhomogenous perturbations. 
 In any cases we should have at least two different types of matter to get a static solution (whatever
stable or unstable) except for a very special case of $w$ exactly equals to $-1/3$.

On the contrary, more complicated structure of equations of motion in modified gravity allows us to get static solutions
with only one matter content of the Universe. This possibility have been already remarked in $f(T)$ gravity, and it is shown
that for some range of $w$ it is possible to get stable static solutions \cite{Wu:2011xa}, \cite{Li:2013xea}  \cite{Skugoreva_2020}.  

Analogs of Einstein static Universe (with matter {\it and}
cosmological constant) have been considered previously  in the context of $f(R)$  \cite{PhysRevD.76.084005}, \cite{PhysRevD.78.044011} \cite{Goheer_2009}, with the $R+R^2$ case included. 

The structure of our paper is as follows. In Sec.2 we remind a reader the conditions of existence and stability of a static
solutions in GR. In Sec.3 we write down the static solution in $R+R^2+R_{ab}R^{ab}$ type of gravity for a one-component isotropic Universe
 which can be considered as subclass of the studied in \cite{PhysRevD.76.084005}. Stability of static solution in quadratic gravity
 with respect to isotropic perturbation 
 is addressed in Sec.4. The static isotropic open case appears to be unstable, while the closed Universe has much simpler conditions for stability as compared to the two-component case shown in \cite{PhysRevD.76.084005}. In Sec.5 
we generalize the closed Universe to the non isotropic case to Bianchi IX and specifically check the size of the oscillatory attractor. 
Finally, Sec.6 contains a brief summary of our results.

\section{Static solutions in GR}
In GR it is possible to find cosmological static solutions in two-fluid systems.
If we consider the metric of an isotropic spatially curved Universe
\begin{align}
ds^{2}=-dt^{2}+\mathcal{R}^2\left(\frac{dr^{2}}{1-kr^{2}}+r^{2}(d\theta^{2}+\sin(\theta)^{2}d\phi^{2})\right)\label{l-e}
    \end{align}
filled with two types of fluid with positive energy densities $\rho_1$ and $\rho_2$ 
and the equation of state parameters $w_1$ and $w_2$ respectively, the equation for cosmic acceleration reads
\begin{equation}
    \frac{\ddot{\mathcal{R}}}{\mathcal{R}}=-\frac{4 \pi G}{3} \rho_1(1+3w_1)-\frac{4 \pi G}{3} \rho_2 (1+3w_2).
\end{equation}
From this equation it is clear that for positive energy densities zero acceleration can be got if one fluid has $w_1<1/3$
and the second fluid has $w_2>-1/3$, so violation of Strong Energy Condition is required. Only with an exceptional case of $w=-1/3$ exactly, this 
equation can be satisfied for a Universe with only one perfect fluid. For the two-fluid 
system the condition for zero acceleration in terms of densities can be written as 
\begin{equation}
    \rho_2=-\rho_1 \frac{1+3w_1}{1+3w_2}
    \label{GR}
\end{equation}
The Friedmann equation
\begin{equation}
    H^2+\frac{k}{\mathcal{R}^2}=\frac{8 \pi G}{3} (\rho_1+\rho_2)
\end{equation}
shows that positive energy density condition requires positive spatial curvature
for the static solution to exist. Setting $H=0$ and using \eqref{GR} we get the expression 
connecting the scale factor and matter energy density.

The classical Einstein static Universe corresponds to $w_1=-1$ (a cosmological constant) and
$w_2=0$ (a dust matter). Simple stability analysis shows that any solution with positive energy
densities is unstable independently of the particular values of $w_1$ and $w_2$.

If we consider a (rather unphysical) situation when one of the perfect 
fluid considered has a negative energy density, the solution can be stable in some situations.
For its existence it requires either two types of matter with $w_{1,2}<-1/3$ or with $w_{1,2}>-1/3$. In the former case the condition 
for stability is $w_2<w_1$ where $w_1$ corresponds to positive energy matter. In the latter case the condition has
the opposite form $w_2>w_1$.
Strictly speaking, negative energy density is dangerous for a physical theory not {\it per se}, but when it can reach
an energy unbounded from below. From this point of view there exists one particular interesting example of a {\it stable} static 
solution in GR, realising in a Universe filled by a positive density matter with $w<-1/3$ and a negative cosmological
constant.

\section{Isotropic static solutions in quadratic gravity \label{static_solutions}}
We consider general quadratic gravity
\begin{align}
L_g=\frac{1}{16G\pi}\left\{\beta R^2 +\alpha \left(R_{ab}R^{ab} -\frac{1}{3} R^2\right)+R\right\}.
\label{acao}
\end{align}
Metric variations result in the field equations
\begin{equation}
G_{ab}+\left(\beta-\frac{1}{3}\alpha\right)H_{\: ab}^{(1)}+\alpha H_{\: ab}^{(2)}=8\pi GT_{ab},\label{eq.campo}
\end{equation}
where 
\begin{eqnarray*}
 &  & G_{ab}=R_{ab}-\frac{1}{2}g_{ab}R,\\
 &  & H_{ab}^{(1)}=-\frac{1}{2}g_{ab}R^{2}+2RR_{ab}+2g_{ab}\square R-2\nabla_a\nabla_bR,\\
 &  & H_{ab}^{(2)}=-\frac{1}{2}g_{ab}R^{cd}R_{cd}+\square R_{ab}+\frac{1}{2}g_{ab}\square R-\nabla_a\nabla_bR\\
 &  & +2R^{cd}R_{cbda}
\end{eqnarray*}
and the perfect fluid classical source is $T_{ab}=(\rho+p)u_au_b+pg_{ab}$. The time like vector $u^a$ is geodesic and vorticity free. 

It is possible to find isotropic static solutions for negatively and positively spatially curved Universes with a one component classical source $\rho$ as follows. 

The isotropic FLRW line element \eqref{l-e} is chosen with scale factor $\mathcal{R}=e^a$, so 
covariant conservation of the source imposes the well known density evolution 
\[
\rho=\frac{\rho_{0}}{e^{3a(1+w)}}.
\]
Since the dynamics is isotropic, the dynamical equation have only terms originating  from  the Einstein-Hilbert part of the
action \eqref{acao} and the part containing the coupling constant $\beta$. Obtained by a standard way, the isotropic part
of the equations of motion are
\begin{align}
2\beta\dddot{a}\dot{a}+6\beta\ddot{a}\dot{a}^{2}-\beta\ddot{a}^{2}+\frac{k^{2}\beta}{e^{4a}}+\frac{1}{6}\dot{a}^{2}-\frac{4\pi G\rho_{0}}{9e^{3a(1+w)}}+k\left(-2\beta\frac{\dot{a}^{2}}{e^{2a}}+\frac{1}{6e^{2a}}\right)=0
\label{eq-00}
\end{align}
and the $11$ equation which contains the dynamic
\begin{align}
 & -12\beta\dddot{a}\dot{a}-18\beta\ddot{a}\dot{a}^{2}-2\beta\ddddot{a}-9\beta\ddot{a}^{2}+\frac{\beta k^{2}}{e^{4a}}\nonumber\\
 & -\frac{1}{3}\ddot{a}-\frac{1}{2}\dot{a}^{2}-\frac{4\pi Gw\rho_{0}}{3e^{3a(1+3w)}}+k\left(\frac{4\beta\ddot{a}}{e^{2a}}+\frac{2\beta\dot{a}^{2}}{e^{2a}}-\frac{1}{6e^{2a}}\right)=0.
 \label{eq-11}
\end{align}
 For the spatial positively and negatively curved cases, $k=\pm1,$
if all derivatives of $a$ are zero we have the following static solution
\begin{eqnarray}
 && a_s=-\frac{1}{2}\ln\left(-\frac{1+3w}{6\beta k\left(-1+3w\right)}\right)\label{a_est}\\
 && \rho_s=1/8\,{\frac {1+3\,w}{\pi \,G\beta\, \left( -1+3\,w \right) ^{2}}}.
\end{eqnarray}

This solution is real for the positively curved case $k=1$ if $\beta>0$
and $-1/3<w<1/3$ while when $\beta<0$ the solution is real for $-1<w<-1/3$
and $1/3<w<1.$ For the negatively curved case $k=-1$ situation is
reversed: when $\beta>0$ the solution is real for $-1<w<-1/3$ and
$1/3<w<1$ and when $\beta<0$ for $-1/3<w<1/3$. 

\section{Stability with respect to isotropic perturbations \label{isotropic_spherical}}
In this section we consider the stability of the solutions written down in the section \ref{static_solutions} for isotropic perturbations. Linearizing these equations \eqref{eq-00} and \eqref{eq-11} near the static solution 
 we, 
independently of the spatial
curvature being positive or negative  have $4$
eigenvalues 
\begin{equation}
\lambda=\pm\frac{i\sqrt{3}\sqrt{9w+1\pm\sqrt{117w^{2}+6w-3+108w^{3}}}}{6\sqrt{\beta}\sqrt{-1+3w}}.
\label{lambdas_isotropic}
\end{equation}

As we mention in section \ref{static_solutions}, stability does not depend on $\alpha$ for the pure isotropic modes. There are two intervals in which all frequencies are real. Remind
that real frequencies correspond to pure imaginary eigenvalues. For
$\beta>0$ and $\beta<0$ we have respectively 
\[
-1<w<-(1+\sqrt{17})/24\;\;\mbox{and}\;\;(-1+\sqrt{17})/24<w<1/3.
\]

So it is possible to obtain stable static positively curved
$k=1$ universe with positive energy density choosing the EOS parameter
in the range $-1/3<w<-(1+\sqrt{17})/24$ as long as $\beta>0.$ Note, that such a matter does not violate
the Strong Energy Condition. Numerically, this is a rather
narrow interval of the EOS parameter from $\sim -0.33$ to $\sim -0.21$. 
We should  also remark that 
the range for existence of static solution with $\beta<0$ does not overlap with 
the stability interval for $w$, so there are no stable static solutions with $\beta<0$.
From physical point of view, the case of $\beta>0$ is more interesting since it includes the possibility of inflation scenario. 

It is also possible to obtain static stable negatively curved
universes $k=-1$ with the EOS parameter in the range $-1<w<-1/3$
for $\beta>0$ and for $\beta<0$ in the range $(-1+\sqrt{17})/24<w<1/3$.
For negative spatial curvature in both cases the energy density is
negative. 
In a physically preferred situation  with $\rho>0$ and $\beta>0$
the static solution
with $k=-1$ is always  unstable and requires $1/3<w<1$.
Here it must be emphasized that in standard GR it is not even possible
a static spatial negatively curved universe with positive energy density.

  \begin{figure}[htpb]
 \begin{center}
\begin{tabular}{c }
\resizebox{\halfsize}{!}{\includegraphics{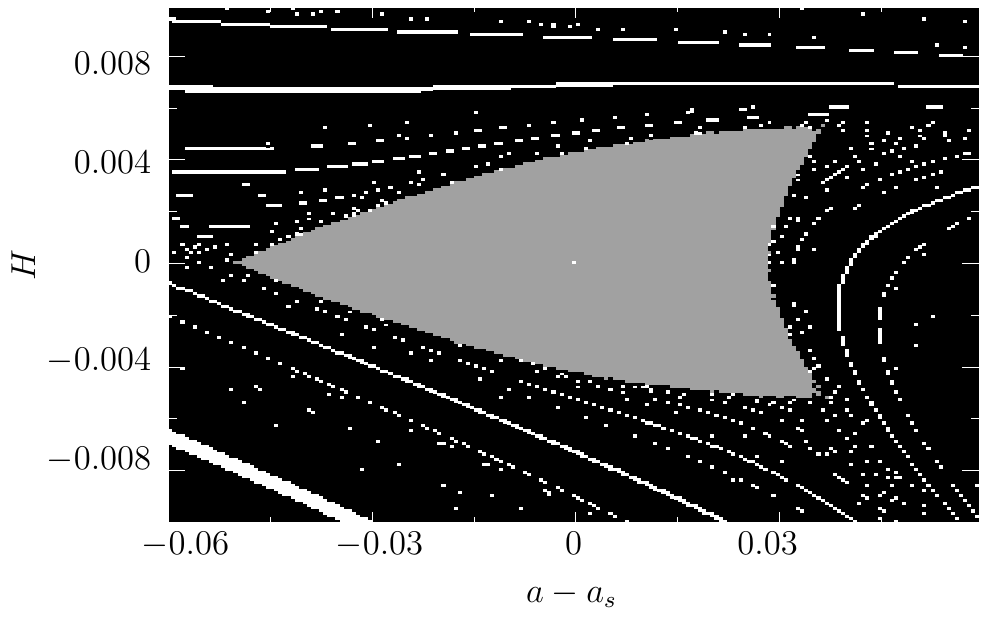}}  
   \end{tabular}
    \end{center}
    \caption{We choose $\beta=10$ with EOS parameter in the stable region $w=-0.22$ for a spatially positively curved universe $k=1$. The x axis marks the difference between initial $a$ and the value for $a_s=2.839985913$ for the static solution given in (\ref{a_est}) and in the y axis is the initial value for $H=\dot{a}$. This section is made for initial $\dot{H}=0$. Black points refer to the singularity attractor, gray points is the attractor to the stable oscillatory region, and the set of white points is the attractor to the asymptotic scalaron behavior $e^{i\omega t}$ with $\omega=1/\sqrt{6\beta}$.  The static solution is located in the 
    center of the plot and is surrounded by the stable oscillatory region. \label{fig1}}
    \end{figure}
\begin{figure}[h]
 \begin{center}
\begin{tabular}{c} 
\resizebox{\halfsize}{!}{\includegraphics{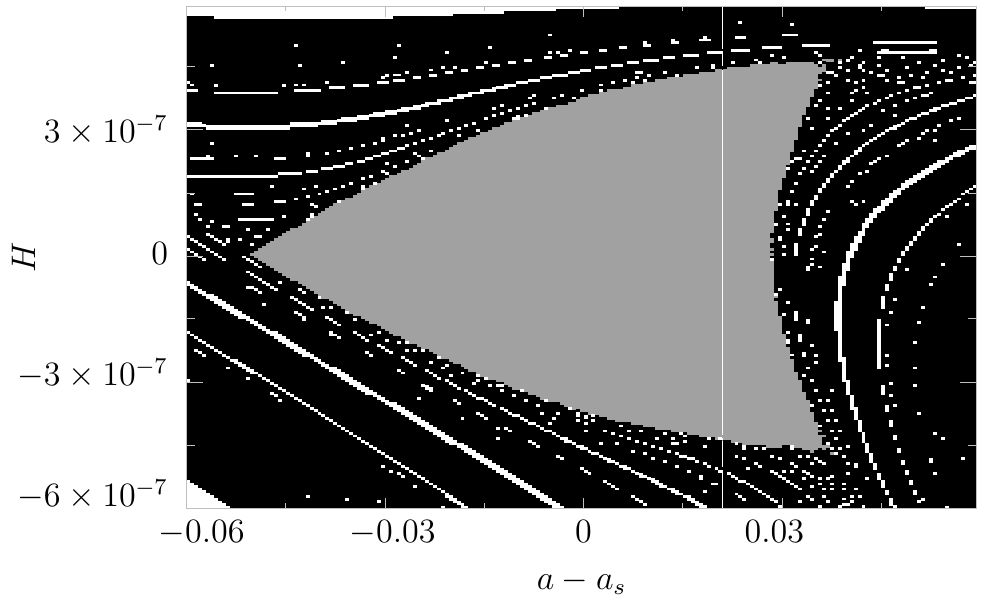}}  
   \end{tabular}
    \end{center}
    \caption{Same plot as Figure \ref{fig1} for $\beta=1.305\times 10^9$ which is the value set by CMBR observations. For this $\beta$ the value for the logarithmic of scale factor for static solution is $a_s=12.18342780$. In this plot grey points correspond to initial conditions which oscillate near static solution. Black points are initial conditions which reach the singularity and as in Figure \ref{fig1} white points reach asymptotic scalaron behavior. We specifically checked that there is no initial condition resulting in appropriate inflation with $\sim 60$ e-folds in this state space. It is also shown that the stripe of the stable oscillatory region is narrowed in the $H$ direction when $\beta$ increases.\label{fig4}}
    \end{figure}

    We choosed $\beta=10$ just for qualitative behavior unless in Figure \ref{fig4} and \ref{fig5}. In Figure \ref{fig1} it is shown the basin of the stable static orbit for EOS parameter chosen in the stable region $w=-0.22$. It is addressed the  positively curved $k=1$ universe and the x axis marks the difference between initial $a$ and the value $a_s=2.839985913$ for the static solution given in (\ref{a_est}) and the y axis is the initial value for $H=\dot{a}$. All initial conditions are set with $\dot{H}=0$.
    The stability region is marked by gray color. Outside of stability region a trajectory either go directly to singularity (black zone) or experience prolonged scalaron oscillations, possibly preceded by 
    inflationary regime   $H=-(t-t_0)/(36\beta)$  (white zone).

     In Figure \ref{fig4} $w=-0.22$ and also $k=1$ as in Figure \ref{fig1} now with $\beta=1.305\times 10^9$ which is the value set by CMBR observations \cite{Ade:2015lrj}. We can see that the stability region is shrinked
     considerably in vertical dimension. As for white zone,
     it was specifically checked that there are no initial conditions near the static solution that converge to the inflationary solution $H=-(t-t_0)/(36\beta)$ with the required number of $\sim 60$ e-folds,
     which is natural since the boundary value of Hubble parameter is very small for this $\beta$. This means
     that realistic trajectories in Starobinsky inflation scenario can not originate near the static solution.
     
    From these plots we can see also that trajectories starting outside the  stability region but close to it can reach the asymptotic scalaron regime leading to significant growth of scale factor, in contrast to $f(T)$ gravity with the same matter content, where trajectories starting close to stability zone 
    are separated from large scale factor regions due to nonstandard singularities \cite{Skugoreva_2020}. Nonstandard singularities are typical for extensions of GR with second order dynamical equations and are 
    absent in such fourth order theory as the quadratic gravity.
\section{Stability with respect to anisotropic perturbations\label{shear}}
In this section stability in the positively curved case is explored with respect to anisotropic perturbations, so we need 
the full form of equations of motion for the general quadratic lagrangian \eqref{acao} given in \eqref{eq.campo}. Tetrad base and proper time is chosen so that the metric is 
\begin{align}
g=\left(   
\begin{array}{cccc}
-1 & 0 &0 & 0\\
0 & 1 & 0 & 0\\
0 & 0 & 1& 0\\
0 & 0 & 0 & 1
\end{array}
\right)
\label{metric}
\end{align}
Covariant conservation of the source implies 
\[
\rho=\frac{\rho_0}{\exp\left[ \int 3(w +1)Hdt \right]}. 
\]
The vector $u=(1,0,0,0)$ is geodesic and vorticity free
\begin{eqnarray}
&&\nabla_0u_i=\nabla_iu_0=\nabla_0u_0=0\nonumber\\
&&\nabla_iu_j=H\delta_{ij}+\sigma_{ij},\;\; \mbox{i,i=1,2,3},
\label{t_gamma1}
\end{eqnarray}
while shear is chosen as
\begin{align}
\sigma_{ij}=\left(   
\begin{array}{ccc}
-2\sigma_+ & 0&0\\
0 & \sigma_+ +\sqrt{3}\sigma_- & 0 \\
0 & 0 & \sigma_+-\sqrt{3}\sigma_- .
\end{array}
\right)
\label{sigma}
\end{align}
Zero shear corresponds to the isotropic cases discussed both in sections \ref{static_solutions} and \ref{isotropic_spherical}. Since there's shear, the scale factor does not exist and  a ``mean scale factor'' must be defined $e^{\int Hdt }$. Also, it is chosen as the variable instead of the scale factor, the logarithmic of scale factor which is called $a$ as follows
\begin{equation}
a=\int Hdt. \label{log_a}
\end{equation}
Since the basis vectors $e_a$ are not rigid, the connection is uniquely determined by metricity and zero torsion, respectively 
\begin{align*}
&\nabla_a g_{bc}=0\\
&\nabla_a e_b-\nabla_b e_a=[e_a,e_b],
\end{align*}
where $g_{ab}$ is given by \eqref{metric}. Since the interest is in investigating only spatially homogeneous models, the commutator of the spatial part of the basis is 
\begin{align*}
[e_i,e_j]=-C^k_{ij}e_k,\;\;\;i,j,k=1,2,3. 
\end{align*}
We decided to investigate in this article the stability of the static solution in the presence of shear and for this reason it is necessary to introduce appropriate structure constants for Bianchi IX which is the anisotropic generalization of the Friedmann closed model 
\begin{align*}
C^1_{23}=2d(t) && C^2_{31}=2b(t) && C^3_{12}=2c(t).
\end{align*}
These structure constants are the appropriate ones that satisfy the  Maurer-Cartan equation 
\begin{align*}
d\omega^k=\frac{1}{2}C^k_{ij}\omega^i\wedge \omega^j,
\end{align*}
where the $\omega^i$ are the left invariant 1-form basis. The spatial part of the connection is 
\begin{align*}
\Gamma_{ijk}=\frac{1}{2}\left( C_{ikj} -C_{jki}-C_{kji}\right),
\end{align*}
with non zero components 
\begin{align}
&\Gamma_{123}=d(t)+b(t)-c(t) & \Gamma_{132}=-d(t)-c(t)+b(t)\nonumber \\
&\Gamma_{312}=c(t)+d(t)-b(t) & \Gamma_{321}=-c(t)-b(t)+d(t)\nonumber\\
&\Gamma_{213}=-b(t)-d(t)+c(t) & \Gamma_{231}=b(t)+c(t)-d(t),
\label{sp_gamma}
\end{align}
while $\nabla_ag_{bc}=0$ implies for \eqref{metric} that $\Gamma_{abc}=-\Gamma_{bac}$.

This choice of structure constant for the group results in the following non zero 3-curvature components of the Riemann tensor ${}^3R^a_{bcd}$ which are constant in the $t=const.$ slices
\begin{align*}
&{}^3R^3_{131}=-2dc+c^2-3b^2+d^2+2cb+2db & {}^3R^3_{232}=-2cb+c^2-3d^2+b^2+2dc+2db\\
&{}^3R^2_{121}=d^2-2db-3c^2+b^2+2cb+2dc
\end{align*}
In this setting remind that we are choosing as variable the connection instead of the metric and following \cite{wainwright2005dynamical} there's an additional condition which guaranties that the correct choice is made which is Jacobi identity $R_{abcd}+R_{acdb}+R_{adbc}=0$. Jacobi identity results in the following first order differential equations which must be satisfied together with the field equations
\begin{align}
&-\dot{d}-(4\sigma_++H)d=0 &-\dot{b}-(H-2\sigma_+-2\sqrt{3}\sigma_-)b=0\nonumber\\
&\dot{c}+(2\sqrt{3}\sigma_-+H-2\sigma_+)c=0.\label{eqs_curvature}
\end{align}
This equations \eqref{eqs_curvature} together with \eqref{sp_gamma}, \eqref{t_gamma1} and \eqref{sigma} completely specify the connection and the general field equations \eqref{eq.campo}  for this situation with perfect fluid source are written in the Appendix. 

For zero shear all the above relations converge to the inverse scale factor and components for the Einstein tensor, for instance, are the same as for the closed Friedmann model
\begin{align*}
& b=c=d=1/e^a=\exp \left(-\int Hdt\right), & G_{00}=3H^2+3e^{-2a}, \hspace{0.5cm} G_{ii}=-2\dot{H}-3H^2-e^{-2a}.
\end{align*}

\begin{figure}[htpb]
 \begin{center}
\begin{tabular}{c } 
\resizebox{\halfsize}{!}{\includegraphics{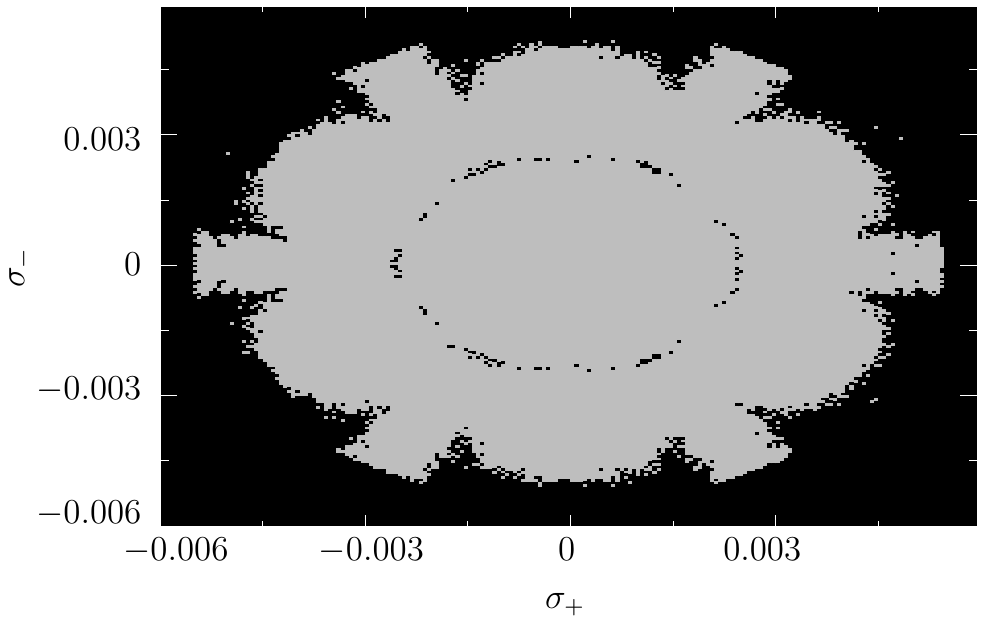}}  
   \end{tabular}
    \end{center}
    \caption{A mesh of initial conditions in $\sigma_+$, versus $\sigma_-$. This plot was made to check the the attractor to the stable region when $\alpha>0$ given in \eqref{a_ma_0} with $\beta=1$, $\alpha=10$ and $w=-0.22$ with static $a_s=1.688693366$. Gray points are initial conditions which oscillate near the static solution, and black points approach the singularity. As expected there are no initial conditions which asymptotically approach the scalaron since for $\alpha>0$ the Minkowski weak field limit becomes unstable. \label{fig5}}
    \end{figure} 
  
Besides the eigenvalues which are present in the isotropic case \eqref{lambdas_isotropic} there are these additional eigenvalues 
\begin{eqnarray}
& & \lambda=0,\\
& & \lambda=\pm i/6\,\sqrt {6}\sqrt {{\frac {-30\,\alpha\,w+9\,\beta\,w+9\,\beta-10\,
\alpha\pm 3\,\sqrt\Delta}{ \left( 3\,w-1
 \right) \alpha\,\beta}}}\label{add_freq}
\end{eqnarray}
where 
\[
\Delta= {36\,{\alpha}^{2}{w}^{2}-12\,\alpha\,{w}^{2}\beta-16\,
\alpha\,w\beta+24\,{\alpha}^{2}w+9\,{\beta}^{2}{w}^{2}+18\,{\beta}^{2}
w+9\,{\beta}^{2}-4\,\alpha\,\beta+4\,{\alpha}^{2}}.
\]

\begin{figure}[htpb]
 \begin{center}
\begin{tabular}{c } 
 \resizebox{\halfsize}{!}{\includegraphics{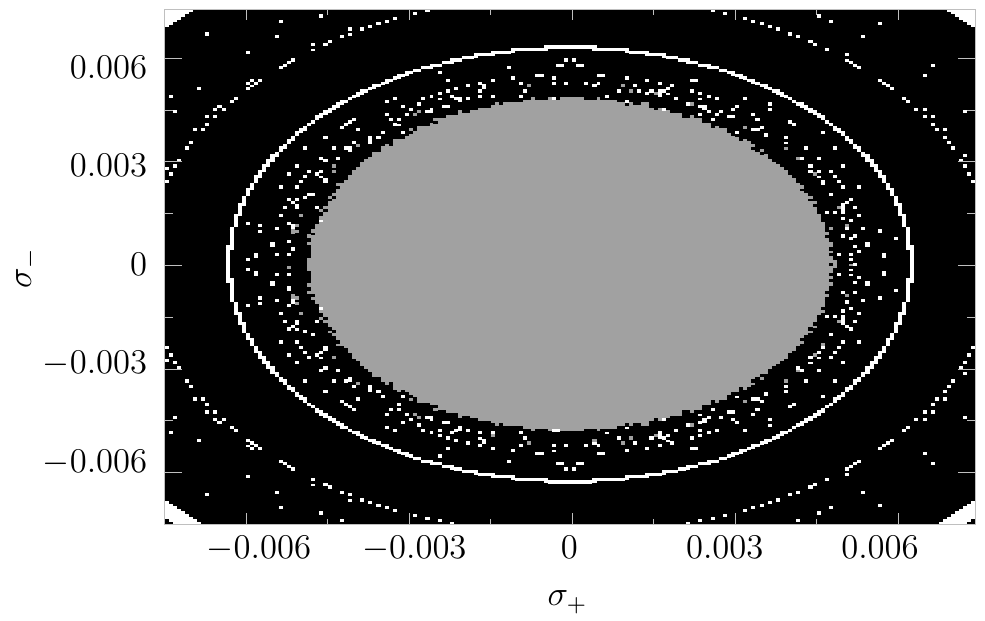}}  
   \end{tabular}
    \end{center}
    \caption{In this mesh it is verified the stability of the static solution shown in Figure \ref{fig1} with respect to shear. For this $\beta=10$, $\alpha=-1$ and $w=-0.22$ with static $a_s=2.839985913$ as in Figure \ref{fig1}, and $\sigma_+$ and $\sigma_-$ are the only non zero initial conditions. Again, gray points oscillate near the static solution, black points are initial conditions that hit the singularity, and white points are initial conditions that asymptote the scalaron oscillations. \label{fig6}}
    \end{figure}


Note that in this section $\rho$ was not excluded from the constraint equation. This formally leads to zero corresponding eigenvalue, since for any $\rho$ is possible to find a static solution. That is why the zero eigenvalue does not affect stability. 

For $\beta<0$ as we saw in section \ref{isotropic_spherical}, stability is not compatible with the existence of the static solution.

Now we turn to positive $\beta>0$. In this case, the stable range of $w$ is given by $-(1/3)(4\alpha-9\beta)/(4\alpha-3\beta)<w<1/3$ for the positive sign $+3\sqrt{\Delta}$ in \eqref{add_freq}. While for the negative sign $-3\sqrt{\Delta}$ in \eqref{add_freq} this stable range is $-1/3<w<1/3$. Superposition of the stability interval of the pure isotropic modes from section \ref{isotropic_spherical}, gives the stable intervals
\begin{align}
&-(1/3)(4\alpha-9\beta)/(4\alpha-3\beta)<w<-(1+\sqrt{17})/24&&\mbox{for } \alpha>0,&& 0<\beta<\frac{4\alpha}{3}\left(\frac {7-\sqrt {17}}{23-\sqrt {17}}\right) \label{a_ma_0}\\
&-1/3<w<-(1+\sqrt{17})/24 && \mbox{for } \alpha<0\label{a_me_0}
\end{align}
 
We can see that for $\alpha<0$ anisotropic perturbations do not change the stability conditions in comparison with the
isotropic case. On the contrary, for $\alpha>0$ the zone of stability is narrower from the viewpoint of possible values of $w$.
Moreover, for 
\begin{align*}
\beta>\frac{4\alpha}{3}\left(\frac {7-\sqrt {17}}{23-\sqrt {17}}\right)
\end{align*}
the static solution is always unstable independently of $w$.

Figure \ref{fig5} shows the situation with $\beta=1$ and
positive $\alpha=10$. The plot in Figure \ref{fig6}  have been made  with the same choice for $\beta=10$, and EOS parameter $w=-0.22$ as in Figure \ref{fig1}, with $\alpha=-1.$ In both cases we can see numerically that stability region exists when shear perturbations are taken into account.

    We also briefly mention the stability of the weak field flat Minkowski space according to quadratic gravity, \eqref{acao}. As already mentioned in the isotropic case there's a scalar degree of freedom with mass $m_s=1/\sqrt{6\beta}$. Besides this one there's also a spin 2 massive ghost with mass $m_2=1/\sqrt{-\alpha}$, see for example \cite{Stelle:1977ry} and \cite{vanDam:1970vg}. Comparing Figures \ref{fig5} and \ref{fig6} it's possible to see that if the spin 2 mode becomes tachyonic with $\alpha>0$ the weak field scalaron basin present in Figures \ref{fig1}, \ref{fig4} and \ref{fig6} is absent in Figure \ref{fig5}. This is expected since the weak field limit becomes unstable for $\alpha>0$.
    
 \section{Conclusions}
 In our paper we pointed out the possibility of static cosmological solution for a Universe filler by only one type of
 perfect fluid in quadratic gravity. In this sense it represent a solution with no direct analog in GR since static solution
 in GR requires at least two different types of matter  -- for example, the cosmological constant and a dust in original
 Einstein solution (apart from a very special case of matter with $w=-1/3$ exactly). Moreover, quadratic gravity allows
 the existence of static solution in negatively curved Universe with positive energy density which is totally impossible
 in GR independently of the number of matter types.
 
 Some of these solutions appear to be stable. Stability requires positivity of spatial constant and positivity of the 
 coupling constant $\beta$. Under this condition a solution is stable with respect to isotropic perturbations if the
 equation of state parameter lies in a rather narrow interval $(-1/3, -(1+\sqrt{17})/24)$.
 
 In the anisotropic case we have two more degrees of freedom which can impose further restrictions on $w$. Our study shows that 
they do not impose any other restrictions on the value of $w$ when the coupling 
 constant $\alpha$ is negative. For positive $\alpha$ the picture of stability is more complicated and depends on 
  the ratio $\alpha/ \beta$. 
 If this ratio is smaller than $(3/4)
 (23-\sqrt{17})/(7-\sqrt{17})$, then the static solution is unstable for any $w$. If
  this ratio exceeds this value, stability conditions  require
 additional restriction on the value of $w$. 
 
 As for static solution with negative spatial curvature, it is unstable for any $w$ of
 its zone of existence.
 
 \section*{Acknowledgments}
 AT is supported by the Russian Government Program of Competitive Growth of Kazan Federal University and RSF grant 21-12-00130. D. M. thanks FAPDF {\it visita t\'ecnica no.}  00193- 00001537/2019-59 for partial support.

\section*{Appendix} 
In this appendix the field equations for \eqref{eq.campo} in presence of shear and perfect fluid source with EOS parameter $w$ are presented. Together with \eqref{eqs_curvature} and
\begin{align*}
    \dot{\rho}=-3H(1+w)\rho
\end{align*}
the dynamical system is completely defined as follows
\begin{align*}
&y_1=H(t)&&y_2=\dot{H}&&y_3=\ddot{H}&&y_4=d(t)&&y_5=b(t)&& y_6=c(t)\\
&y_7=\rho(t)&& y_8=\sigma_+(t)&&y_9=\dot{\sigma}_+&&y_{10}=\ddot{\sigma}_+&&y_{11}=\sigma_-(t)&&
y_{12}=\dot{\sigma}_-&&y_{13}=\ddot{\sigma}_-\\
\end{align*}
{\tiny
\begin{align*}
&\dddot{H}=4/3\,y_{{4}}y_{{6}}\sqrt {3}y_{{12}}-4/3\,y_{{4}}y_{{5}}\sqrt {3}y_{{
12}}-{\frac {32}{3}}\,y_{{8}}{y_{{6}}}^{2}\sqrt {3}y_{{11}}+{\frac {32
}{3}}\,y_{{8}}{y_{{5}}}^{2}\sqrt {3}y_{{11}}+8/3\,\sqrt {3}y_{{11}}y_{
{1}}{y_{{6}}}^{2}-8/3\,\sqrt {3}y_{{11}}y_{{1}}{y_{{5}}}^{2}-4/3\,{y_{
{6}}}^{2}\sqrt {3}y_{{12}}+4/3\,{y_{{5}}}^{2}\sqrt {3}y_{{12}}\\
&+16/3\,y
_{{6}}y_{{1}}y_{{8}}y_{{5}}-8/3\,y_{{6}}y_{{1}}y_{{4}}y_{{8}}-8/3\,y_{
{4}}y_{{1}}y_{{8}}y_{{5}}+16/3\,y_{{8}}y_{{5}}\sqrt {3}y_{{11}}y_{{4}}
-8/3\,\sqrt {3}y_{{11}}y_{{1}}y_{{4}}y_{{6}}+8/3\,\sqrt {3}y_{{11}}y_{
{1}}y_{{4}}y_{{5}}-16/3\,y_{{8}}y_{{6}}\sqrt {3}y_{{11}}y_{{4}}+{
\frac {49}{3}}\,{y_{{11}}}^{2}{y_{{5}}}^{2}\\
&+{\frac {17}{3}}\,{y_{{8}}}
^{2}{y_{{5}}}^{2}+4/3\,y_{{6}}y_{{2}}y_{{5}}+4/3\,y_{{2}}y_{{4}}y_{{5}
}+4/3\,y_{{6}}y_{{2}}y_{{4}}-4\,y_{{1}}y_{{9}}y_{{8}}+4/3\,y_{{6}}y_{{
9}}y_{{4}}+4/3\,y_{{9}}y_{{4}}y_{{5}}-8/3\,y_{{6}}y_{{9}}y_{{5}}-4\,y_
{{11}}y_{{12}}y_{{1}}+16/3\,y_{{1}}{y_{{4}}}^{2}y_{{8}}-{\frac {26}{3}
}\,{y_{{11}}}^{2}y_{{4}}y_{{5}}\\
&+2/3\,{y_{{1}}}^{2}y_{{4}}y_{{5}}-10/3
\,y_{{4}}{y_{{8}}}^{2}y_{{5}}-8/3\,y_{{8}}y_{{1}}{y_{{5}}}^{2}+2/9\,y_
{{4}}y_{{6}}{y_{{5}}}^{2}+2/9\,y_{{6}}{y_{{4}}}^{2}y_{{5}}+2/9\,{y_{{6
}}}^{2}y_{{4}}y_{{5}}-2/3\,{y_{{11}}}^{2}y_{{6}}y_{{5}}+2/3\,y_{{6}}{y
_{{1}}}^{2}y_{{5}}-{\frac {34}{3}}\,y_{{6}}{y_{{8}}}^{2}y_{{5}}-{
\frac {26}{3}}\,{y_{{11}}}^{2}y_{{6}}y_{{4}}\\
&+2/3\,y_{{6}}{y_{{1}}}^{2}
y_{{4}}-10/3\,y_{{4}}y_{{6}}{y_{{8}}}^{2}-8/3\,{y_{{6}}}^{2}y_{{1}}y_{
{8}}+ \frac{\alpha}{\beta}\left( -{\frac {8}{9}}\,y_{{8}}{y_{{6}}}^{2}\sqrt {3}y_{{11}}+{
\frac {8}{9}}\,y_{{8}}{y_{{5}}}^{2}\sqrt {3}y_{{11}}+4/9\,y_{{8}}y_{{5
}}\sqrt {3}y_{{11}}y_{{4}}-4/9\,y_{{8}}y_{{6}}\sqrt {3}y_{{11}}y_{{4}}+{\frac {11}{9}}\,{y_{{11}}}^{2}{y_{{5}}}^{2}\right.\\
&\left.+1/3\,{y_{{8}}}^{2}{y_{{5
}}}^{2}.+1/3\,y_{{1}}y_{{9}}y_{{8}}+1/3\,y_{{11}}y_{{12}}y_{{1}}-4/9\,{
y_{{11}}}^{2}y_{{4}}y_{{5}}+{\frac {4}{27}}\,y_{{4}}y_{{6}}{y_{{5}}}^{
2}+{\frac {4}{27}}\,y_{{6}}{y_{{4}}}^{2}y_{{5}}+{\frac {4}{27}}\,{y_{{
6}}}^{2}y_{{4}}y_{{5}}+2/9\,{y_{{11}}}^{2}y_{{6}}y_{{5}}-2/3\,y_{{6}}{
y_{{8}}}^{2}y_{{5}}-4/9\,{y_{{11}}}^{2}y_{{6}}y_{{4}}\right. \\
&\left. +1/4\,{y_{{1}}}^{
2}{y_{{8}}}^{2}-2\,{y_{{11}}}^{2}{y_{{8}}}^{2}+1/4\,{y_{{11}}}^{2}{y_{
{1}}}^{2}-{\frac {4}{27}}\,{y_{{4}}}^{3}y_{{5}}-{\frac {4}{27}}\,y_{{4
}}{y_{{5}}}^{3}+1/3\,{y_{{6}}}^{2}{y_{{8}}}^{2}+5/3\,{y_{{4}}}^{2}{y_{
{8}}}^{2}+{\frac {11}{9}}\,{y_{{11}}}^{2}{y_{{6}}}^{2}-{\frac {4}{27}}
\,y_{{6}}{y_{{5}}}^{3}-{\frac {4}{27}}\,{y_{{6}}}^{3}y_{{5}}-{\frac {4
}{27}}\,y_{{6}}{y_{{4}}}^{3}\right.\\
&\left. -{\frac {4}{27}}\,{y_{{6}}}^{3}y_{{4}}-1/9
\,{y_{{11}}}^{2}{y_{{4}}}^{2}+1/6\,{y_{{11}}}^{2}y_{{2}}+1/6\,y_{{2}}{
y_{{8}}}^{2}+1/6\,y_{{10}}y_{{8}}+1/6\,y_{{11}}y_{{13}}-{y_{{11}}}^{4}
-1/12\,{y_{{9}}}^{2}-1/12\,{y_{{12}}}^{2}+{\frac {4}{27}}\,{y_{{4}}}^{
4}+{\frac {4}{27}}\,{y_{{5}}}^{4}+{\frac {4}{27}}\,{y_{{6}}}^{4}-{y_{{8}}}^{4} \right)
\end{align*}
\begin{align*}
&+\frac {1}{\beta}\left(-1/4\,{y_{{8}}}^{2}+1/36\,
{y_{{4}}}^{2}-1/4\,{y_{{1}}}^{2}-1/6\,y_{{2}}-1/4\,{y_{{11}}}^{2}-2/3
\,\pi \,Gwy_{{7}}-1/18\,y_{{4}}y_{{5}}+1/36\,{y_{{5}}}^{2}-1/18\,y_{{4
}}y_{{6}}-1/18\,y_{{6}}y_{{5}}+1/36\,{y_{{6}}}^{2}\right)-3\,{y_{{1}
}}^{2}{y_{{8}}}^{2}\\
&-3\,{y_{{11}}}^{2}{y_{{8}}}^{2} -1/3\,{y_{{1}}}^{2}{
y_{{4}}}^{2}+1/3\,{y_{{4}}}^{2}{y_{{5}}}^{2}-3\,{y_{{11}}}^{2}{y_{{1}}
}^{2}-2/9\,{y_{{4}}}^{3}y_{{5}}-2/9\,y_{{4}}{y_{{5}}}^{3}+{\frac {17}{
3}}\,{y_{{6}}}^{2}{y_{{8}}}^{2}+{\frac {65}{3}}\,{y_{{4}}}^{2}{y_{{8}}
}^{2}+{\frac {49}{3}}\,{y_{{11}}}^{2}{y_{{6}}}^{2}-1/3\,{y_{{6}}}^{2}{
y_{{1}}}^{2}-1/3\,{y_{{1}}}^{2}{y_{{5}}}^{2}\\
&+1/3\,{y_{{6}}}^{2}{y_{{5}
}}^{2}-2/9\,y_{{6}}{y_{{5}}}^{3}-2/9\,{y_{{6}}}^{3}y_{{5}}+1/3\,{y_{{6
}}}^{2}{y_{{4}}}^{2}-2/9\,y_{{6}}{y_{{4}}}^{3}-2/9\,{y_{{6}}}^{3}y_{{4
}}+1/3\,{y_{{11}}}^{2}{y_{{4}}}^{2}-9\,y_{{2}}{y_{{1}}}^{2}-2\,{y_{{11
}}}^{2}y_{{2}}-2\,y_{{2}}{y_{{8}}}^{2}-2/3\,{y_{{6}}}^{2}y_{{2}}-2/3\,
y_{{2}}{y_{{5}}}^{2}\\
&-2/3\,y_{{2}}{y_{{4}}}^{2}-6\,y_{{3}}y_{{1}}+4/3\,
{y_{{6}}}^{2}y_{{9}}+4/3\,y_{{9}}{y_{{5}}}^{2}-8/3\,y_{{9}}{y_{{4}}}^{
2}-2\,y_{{10}}y_{{8}}-2\,y_{{11}}y_{{13}}-3/2\,{y_{{11}}}^{4}-9/2\,{y_
{{2}}}^{2}-2\,{y_{{9}}}^{2}-2\,{y_{{12}}}^{2}+1/18\,{y_{{4}}}^{4}+1/18
\,{y_{{5}}}^{4}+1/18\,{y_{{6}}}^{4}\\
&-3/2\,{y_{{8}}}^{4}
\end{align*} 
\begin{align*}
&\dddot{\sigma}_+=8/3\,y_{{4}}y_{{6}}\sqrt {3}y_{{12}}-8/3\,y_{{4}}y_{{5}}\sqrt {3}y_{{12}}+16\,y_{{8}}{y_{{6}}}^{2}\sqrt {3}y_{{11}}-16\,y_{{8}}{y_{{5}}}^{2
}\sqrt {3}y_{{11}}+16/3\,\sqrt {3}y_{{11}}y_{{1}}{y_{{6}}}^{2}-16/3\,
\sqrt {3}y_{{11}}y_{{1}}{y_{{5}}}^{2}-7\,y_{{2}}y_{{8}}y_{{1}}+16\,y_{
{12}}y_{{8}}y_{{11}}\\
&+24\,{y_{{11}}}^{2}y_{{8}}y_{{1}}+ \frac{\beta}{\alpha}\left( 16\,y_{{
8}}{y_{{6}}}^{2}\sqrt {3}y_{{11}}-16\,y_{{8}}{y_{{5}}}^{2}\sqrt {3}y_{
{11}}+84\,y_{{2}}y_{{8}}y_{{1}}+24\,y_{{12}}y_{{8}}y_{{11}}+36\,{y_{{
11}}}^{2}y_{{8}}y_{{1}}+8\,y_{{6}}y_{{1}}y_{{8}}y_{{5}}+8\,y_{{6}}y_{{
1}}y_{{4}}y_{{8}}+8\,y_{{4}}y_{{1}}y_{{8}}y_{{5}}+16\,y_{{8}}y_{{5}}
\sqrt {3}y_{{11}}y_{{4}}\right. \\
&\left.-16\,y_{{8}}y_{{6}}\sqrt {3}y_{{11}}y_{{4}}+8
\,{y_{{11}}}^{2}{y_{{5}}}^{2}-8\,{y_{{8}}}^{2}{y_{{5}}}^{2}-16\,y_{{6}
}y_{{2}}y_{{5}}+8\,y_{{2}}y_{{4}}y_{{5}}+8\,y_{{6}}y_{{2}}y_{{4}}+8\,y
_{{6}}y_{{9}}y_{{4}}+8\,y_{{9}}y_{{4}}y_{{5}}+8\,y_{{6}}y_{{9}}y_{{5}}
-4\,y_{{1}}{y_{{4}}}^{2}y_{{8}}+8\,{y_{{11}}}^{2}y_{{4}}y_{{5}}+16\,{y
_{{1}}}^{2}y_{{4}}y_{{5}}\right.\\
&\left.-8\,y_{{4}}{y_{{8}}}^{2}y_{{5}}-4\,y_{{8}}y_{
{1}}{y_{{5}}}^{2}-8/3\,y_{{4}}y_{{6}}{y_{{5}}}^{2}+16/3\,y_{{6}}{y_{{4
}}}^{2}y_{{5}}-8/3\,{y_{{6}}}^{2}y_{{4}}y_{{5}}-16\,{y_{{11}}}^{2}y_{{
6}}y_{{5}}-32\,y_{{6}}{y_{{1}}}^{2}y_{{5}}+16\,y_{{6}}{y_{{8}}}^{2}y_{
{5}}+8\,{y_{{11}}}^{2}y_{{6}}y_{{4}}+16\,y_{{6}}{y_{{1}}}^{2}y_{{4}}-8
\,y_{{4}}y_{{6}}{y_{{8}}}^{2}\right. \\
&\left. -4\,{y_{{6}}}^{2}y_{{1}}y_{{8}}+72\,y_{{8
}}{y_{{1}}}^{3}+36\,{y_{{8}}}^{3}y_{{1}}+12\,y_{{8}}y_{{3}}+24\,y_{{9}
}{y_{{1}}}^{2}+12\,{y_{{11}}}^{2}y_{{9}}+36\,y_{{9}}{y_{{8}}}^{2}+12\,
y_{{9}}y_{{2}}-32\,{y_{{1}}}^{2}{y_{{4}}}^{2}+8\,{y_{{4}}}^{2}{y_{{5}}
}^{2}-{\frac {40}{3}}\,{y_{{4}}}^{3}y_{{5}}+8/3\,y_{{4}}{y_{{5}}}^{3}-
8\,{y_{{6}}}^{2}{y_{{8}}}^{2}\right.\\
&\left.+16\,{y_{{4}}}^{2}{y_{{8}}}^{2}+8\,{y_{{
11}}}^{2}{y_{{6}}}^{2}+16\,{y_{{6}}}^{2}{y_{{1}}}^{2}+16\,{y_{{1}}}^{2
}{y_{{5}}}^{2}-16\,{y_{{6}}}^{2}{y_{{5}}}^{2}+{\frac {32}{3}}\,y_{{6}}
{y_{{5}}}^{3}+{\frac {32}{3}}\,{y_{{6}}}^{3}y_{{5}}+8\,{y_{{6}}}^{2}{y
_{{4}}}^{2}-{\frac {40}{3}}\,y_{{6}}{y_{{4}}}^{3}+8/3\,{y_{{6}}}^{3}y_
{{4}}-16\,{y_{{11}}}^{2}{y_{{4}}}^{2}+8\,{y_{{6}}}^{2}y_{{2}}\right.\\
&\left.+8\,y_{{2
}}{y_{{5}}}^{2}-16\,y_{{2}}{y_{{4}}}^{2}-4\,{y_{{6}}}^{2}y_{{9}}-4\,y_
{{9}}{y_{{5}}}^{2}-4\,y_{{9}}{y_{{4}}}^{2}+16/3\,{y_{{4}}}^{4}-8/3\,{y
_{{5}}}^{4}-8/3\,{y_{{6}}}^{4} \right) +16/3\,{y_{{6
}}}^{2}\sqrt {3}y_{{12}}-16/3\,{y_{{5}}}^{2}\sqrt {3}y_{{12}}+8\,y_{{6
}}y_{{1}}y_{{8}}y_{{5}}+8/3\,\sqrt {3}y_{{11}}y_{{1}}y_{{4}}y_{{6}}\\
&-8/
3\,\sqrt {3}y_{{11}}y_{{1}}y_{{4}}y_{{5}}-{\frac {104}{3}}\,{y_{{11}}}
^{2}{y_{{5}}}^{2}-8\,{y_{{8}}}^{2}{y_{{5}}}^{2}+8\,y_{{6}}y_{{9}}y_{{5
}}-20\,y_{{1}}{y_{{4}}}^{2}y_{{8}}-{\frac {32}{3}}\,{y_{{11}}}^{2}y_{{
4}}y_{{5}}-4\,y_{{8}}y_{{1}}{y_{{5}}}^{2}-{\frac {16}{9}}\,y_{{4}}y_{{
6}}{y_{{5}}}^{2}+{\frac {32}{9}}\,y_{{6}}{y_{{4}}}^{2}y_{{5}}-{\frac {
16}{9}}\,{y_{{6}}}^{2}y_{{4}}y_{{5}}\\
&+16/3\,{y_{{11}}}^{2}y_{{6}}y_{{5}
}+16\,y_{{6}}{y_{{8}}}^{2}y_{{5}}-{\frac {32}{3}}\,{y_{{11}}}^{2}y_{{6
}}y_{{4}}-4\,{y_{{6}}}^{2}y_{{1}}y_{{8}}-6\,y_{{8}}{y_{{1}}}^{3}+24\,{
y_{{8}}}^{3}y_{{1}}-y_{{8}}y_{{3}}-11\,y_{{9}}{y_{{1}}}^{2}+8\,{y_{{11
}}}^{2}y_{{9}}+24\,y_{{9}}{y_{{8}}}^{2}-4\,y_{{9}}y_{{2}}-6\,y_{{1}}y_
{{10}}+\frac{1}{\alpha} \left(2/3\,y_{{6}}^{2}\right.\\
&\left. -4/3\,y_{{6}}y_{{5}}+2/3\,y_{{4}}y_{{
6}}-4/3\,{y_{{4}}}^{2}+2/3\,{y_{{5}}}^{2}+3\,y_{{8}}y_{{1}}+y_{{9}}+2/
3\,y_{{4}}y_{{5}}\right)-{\frac {80}{9}}\,{y_{{4}}}^{3}y_{{5}}+{
\frac {16}{9}}\,y_{{4}}{y_{{5}}}^{3}-8\,{y_{{6}}}^{2}{y_{{8}}}^{2}+80
\,{y_{{4}}}^{2}{y_{{8}}}^{2}-{\frac {104}{3}}\,{y_{{11}}}^{2}{y_{{6}}}
^{2}+{\frac {64}{9}}\,y_{{6}}{y_{{5}}}^{3}\\
&+{\frac {64}{9}}\,{y_{{6}}}^
{3}y_{{5}}-{\frac {80}{9}}\,y_{{6}}{y_{{4}}}^{3}+{\frac {16}{9}}\,{y_{
{6}}}^{3}y_{{4}}+16/3\,{y_{{11}}}^{2}{y_{{4}}}^{2}-4\,{y_{{6}}}^{2}y_{
{9}}-4\,y_{{9}}{y_{{5}}}^{2}-20\,y_{{9}}{y_{{4}}}^{2}+{\frac {128}{9}}
\,{y_{{4}}}^{4}-{\frac {64}{9}}\,{y_{{5}}}^{4}-{\frac {64}{9}}\,{y_{{6
}}}^{4}
\end{align*}
\begin{align*}
&\dddot{\sigma}_-=-1/9\,\frac {\sqrt {3}\beta}{\alpha}\left( -24\,y_{{4}}y_{{6}}\sqrt {3}y_{{12}}-24\,y_{{4}}
y_{{5}}\sqrt {3}y_{{12}}+48\,y_{{8}}{y_{{6}}}^{2}\sqrt {3}y_{{11}}+48
\,y_{{8}}{y_{{5}}}^{2}\sqrt {3}y_{{11}}+12\,\sqrt {3}y_{{11}}y_{{1}}{y
_{{6}}}^{2}+12\,\sqrt {3}y_{{11}}y_{{1}}{y_{{5}}}^{2}+12\,{y_{{6}}}^{2
}\sqrt {3}y_{{12}}+12\,{y_{{5}}}^{2}\sqrt {3}y_{{12}}\right.\\
&\left.-108\,{y_{{11}}}^
{3}\sqrt {3}y_{{1}}-216\,\sqrt {3}y_{{11}}{y_{{1}}}^{3}-36\,\sqrt {3}y
_{{11}}y_{{3}}+12\,{y_{{4}}}^{2}\sqrt {3}y_{{12}}-36\,\sqrt {3}y_{{12}
}y_{{2}}-36\,\sqrt {3}y_{{12}}{y_{{8}}}^{2}-72\,\sqrt {3}y_{{12}}{y_{{
1}}}^{2}-108\,{y_{{11}}}^{2}\sqrt {3}y_{{12}}-96\,y_{{8}}y_{{5}}\sqrt 
{3}y_{{11}}y_{{6}}\right.\\
&\left.-24\,\sqrt {3}y_{{11}}y_{{1}}y_{{6}}y_{{5}}+48\,y_{{
8}}y_{{5}}\sqrt {3}y_{{11}}y_{{4}}-24\,\sqrt {3}y_{{11}}y_{{1}}y_{{4}}
y_{{6}}-24\,\sqrt {3}y_{{11}}y_{{1}}y_{{4}}y_{{5}}+48\,y_{{8}}y_{{6}}
\sqrt {3}y_{{11}}y_{{4}}+72\,{y_{{11}}}^{2}{y_{{5}}}^{2}-72\,{y_{{8}}}
^{2}{y_{{5}}}^{2}+72\,y_{{2}}y_{{4}}y_{{5}}-72\,y_{{6}}y_{{2}}y_{{4}}\right.\\
&\left.-
72\,{y_{{11}}}^{2}y_{{4}}y_{{5}}+144\,{y_{{1}}}^{2}y_{{4}}y_{{5}}+72\,
y_{{4}}{y_{{8}}}^{2}y_{{5}}+24\,y_{{4}}y_{{6}}{y_{{5}}}^{2}-24\,{y_{{6
}}}^{2}y_{{4}}y_{{5}}+72\,{y_{{11}}}^{2}y_{{6}}y_{{4}}-144\,y_{{6}}{y_
{{1}}}^{2}y_{{4}}-72\,y_{{4}}y_{{6}}{y_{{8}}}^{2}-252\,y_{{2}}\sqrt {3
}y_{{11}}y_{{1}}-72\,y_{{9}}y_{{8}}\sqrt {3}y_{{11}}\right.\\
&\left.-24\,y_{{6}}y_{{5}
}\sqrt {3}y_{{12}}-108\,{y_{{8}}}^{2}\sqrt {3}y_{{11}}y_{{1}}-96\,y_{{
8}}{y_{{4}}}^{2}\sqrt {3}y_{{11}}+12\,\sqrt {3}y_{{11}}y_{{1}}{y_{{4}}
}^{2}+72\,{y_{{4}}}^{2}{y_{{5}}}^{2}-24\,{y_{{4}}}^{3}y_{{5}}-72\,y_{{
4}}{y_{{5}}}^{3}+72\,{y_{{6}}}^{2}{y_{{8}}}^{2}-72\,{y_{{11}}}^{2}{y_{
{6}}}^{2}+144\,{y_{{6}}}^{2}{y_{{1}}}^{2}\right.\\
&\left.-144\,{y_{{1}}}^{2}{y_{{5}}}^
{2}-48\,y_{{6}}{y_{{5}}}^{3}+48\,{y_{{6}}}^{3}y_{{5}}-72\,{y_{{6}}}^{2
}{y_{{4}}}^{2}+24\,y_{{6}}{y_{{4}}}^{3}+72\,{y_{{6}}}^{3}y_{{4}}+72\,{
y_{{6}}}^{2}y_{{2}}-72\,y_{{2}}{y_{{5}}}^{2}+24\,{y_{{5}}}^{4}-24\,{y_
{{6}}}^{4} \right) -1/9\, \left( -16\,y_{{4}}y
_{{6}}\sqrt {3}y_{{12}}-16\,y_{{4}}y_{{5}}\sqrt {3}y_{{12}}\right.\\
&\left.+176\,y_{{8
}}{y_{{6}}}^{2}\sqrt {3}y_{{11}}+176\,y_{{8}}{y_{{5}}}^{2}\sqrt {3}y_{
{11}}+44\,\sqrt {3}y_{{11}}y_{{1}}{y_{{6}}}^{2}+44\,\sqrt {3}y_{{11}}y
_{{1}}{y_{{5}}}^{2}+18\,y_{{1}}\sqrt {3}y_{{13}}+44\,{y_{{6}}}^{2}
\sqrt {3}y_{{12}}+44\,{y_{{5}}}^{2}\sqrt {3}y_{{12}}-72\,{y_{{11}}}^{3
}\sqrt {3}y_{{1}}+18\,\sqrt {3}y_{{11}}{y_{{1}}}^{3}\right.\\
&\left.+3\,\sqrt {3}y_{{
11}}y_{{3}}-4\,{y_{{4}}}^{2}\sqrt {3}y_{{12}}+12\,\sqrt {3}y_{{12}}y_{
{2}}-24\,\sqrt {3}y_{{12}}{y_{{8}}}^{2}+33\,\sqrt {3}y_{{12}}{y_{{1}}}
^{2}-72\,{y_{{11}}}^{2}\sqrt {3}y_{{12}}+32\,y_{{8}}y_{{5}}\sqrt {3}y_
{{11}}y_{{6}}+8\,\sqrt {3}y_{{11}}y_{{1}}y_{{6}}y_{{5}}-24\,y_{{6}}y_{
{1}}y_{{4}}y_{{8}}+24\,y_{{4}}y_{{1}}y_{{8}}y_{{5}}\right.\\
&\left.+32\,y_{{8}}y_{{5}}
\sqrt {3}y_{{11}}y_{{4}}-16\,\sqrt {3}y_{{11}}y_{{1}}y_{{4}}y_{{6}}-16
\,\sqrt {3}y_{{11}}y_{{1}}y_{{4}}y_{{5}}+32\,y_{{8}}y_{{6}}\sqrt {3}y_
{{11}}y_{{4}}+264\,{y_{{11}}}^{2}{y_{{5}}}^{2}+120\,{y_{{8}}}^{2}{y_{{
5}}}^{2}-24\,y_{{6}}y_{{9}}y_{{4}}+24\,y_{{9}}y_{{4}}y_{{5}}-48\,{y_{{
11}}}^{2}y_{{4}}y_{{5}}\right.\\
&\left.-48\,y_{{4}}{y_{{8}}}^{2}y_{{5}}+48\,y_{{8}}y_{
{1}}{y_{{5}}}^{2}+16\,y_{{4}}y_{{6}}{y_{{5}}}^{2}-16\,{y_{{6}}}^{2}y_{
{4}}y_{{5}}+48\,{y_{{11}}}^{2}y_{{6}}y_{{4}}+48\,y_{{4}}y_{{6}}{y_{{8}
}}^{2}-48\,{y_{{6}}}^{2}y_{{1}}y_{{8}}+21\,y_{{2}}\sqrt {3}y_{{11}}y_{
{1}}-48\,y_{{9}}y_{{8}}\sqrt {3}y_{{11}}+8\,y_{{6}}y_{{5}}\sqrt {3}y_{
{12}}\right.\\
&\left.-72\,{y_{{8}}}^{2}\sqrt {3}y_{{11}}y_{{1}}+32\,y_{{8}}{y_{{4}}}^{
2}\sqrt {3}y_{{11}}-4\,\sqrt {3}y_{{11}}y_{{1}}{y_{{4}}}^{2}-16\,{y_{{
4}}}^{3}y_{{5}}-48\,y_{{4}}{y_{{5}}}^{3}-120\,{y_{{6}}}^{2}{y_{{8}}}^{
2}-264\,{y_{{11}}}^{2}{y_{{6}}}^{2}-32\,y_{{6}}{y_{{5}}}^{3}+32\,{y_{{
6}}}^{3}y_{{5}}+16\,y_{{6}}{y_{{4}}}^{3}+48\,{y_{{6}}}^{3}y_{{4}}\right.\\
&\left.-48\,
{y_{{6}}}^{2}y_{{9}}+48\,y_{{9}}{y_{{5}}}^{2}+64\,{y_{{5}}}^{4}-64\,{y
_{{6}}}^{4} \right) \sqrt {3}-1/9\,{\frac {\sqrt {3}}{\alpha} \left( 6\,y_{{4}}y_{{5}}+6
\,{y_{{6}}}^{2}-6\,y_{{4}}y_{{6}}-9\,\sqrt {3}y_{{11}}y_{{1}}-3\,
\sqrt {3}y_{{12}}-6\,{y_{{5}}}^{2} \right) }
\end{align*}
}
\bibliographystyle{utcaps}
\bibliography{refsR2.bib}

\end{document}